\documentstyle[emulateapj]{article}
\submitted{To appear in ApJ Letters}

\lefthead{Blakeslee et al.}
\righthead{SBF-Density Comparison}

\makeatletter
\newenvironment{inlinetable}{%
\def\@captype{table}%
\noindent\begin{minipage}{0.999\linewidth}\begin{center}\footnotesize}
{\end{center}\end{minipage}\smallskip}

\newenvironment{inlinefigure}{%
\def\@captype{figure}%
\noindent\begin{minipage}{0.999\linewidth}\begin{center}}
{\end{center}\end{minipage}\smallskip}
\makeatother

\newcommand\lta{\mathrel{\rlap{\lower 3pt\hbox{$\mathchar"218$}}
     \raise 2.0pt\hbox{$\mathchar"13C$}}}
\newcommand\gta{\mathrel{\rlap{\lower 3pt\hbox{$\mathchar"218$}}
     \raise 2.0pt\hbox{$\mathchar"13E$}}}
\newcommand\kms{km~s$^{-1}$}
\newcommand\kmsM{km~s$^{-1}\,$Mpc}
\newcommand\etal{{et~al.}} 
\newcommand\sigth{\ifmmode \sigma_{\rm th}\else$\sigma_{\rm th}$\fi}
\newcommand\sigv{\ifmmode \sigma_v\else$\sigma_v$\fi}
\newcommand\mM{\ifmmode(m{-}M)\else$(m{-}M)$\fi}
\newcommand\h{\ifmmode H_0\else$H_0$\fi}
\newcommand\msun{\ifmmode{\hbox{M$_\odot$}}\else{M$_\odot$}\fi}
\newcommand\iras{{\it IRAS}}
\newcommand\bi{\ifmmode \beta_{I}\else$\beta_I$\fi}
\newcommand\bo{\ifmmode \beta_{O}\else$\beta_O$\fi}
\newcommand\czd{$c{z}$--$d$}

\begin{document}

\title{A First Comparison of the SBF Survey Distances with 
the Galaxy Density~Field: Implications for $H_0$ and $\Omega$}

\author{John P. Blakeslee\altaffilmark{1,}\altaffilmark{2},
Marc Davis\altaffilmark{3},
John L. Tonry\altaffilmark{4},\\
Alan Dressler\altaffilmark{5},
and
Edward A. Ajhar\altaffilmark{6}
}
\authoremail{jpb@astro.caltech.edu}
\authoremail{marc@deep.berkeley.edu}
\authoremail{jt@ifa.hawaii.edu}
\authoremail{dressler@omega.ociw.edu}
\authoremail{ajhar@noao.edu}

\altaffiltext{1}{Dept.~of Astronomy, MS 105-24,
California Institute of Technology, Pasadena, CA 91125; jpb@astro.caltech.edu}
\altaffiltext{2}{Also:  Dept.~of Physics,
University of Durham, South Road, Durham, DH1 3LE, England}
\altaffiltext{3}{Dept.~of Astronomy, University of California,
Berkeley, CA~94720}
\altaffiltext{4}{Institute for Astronomy,
University of Hawaii, Honolulu, HI 96822}
\altaffiltext{5}{Carnegie Observatories, 813 Santa Barbara St.,
Pasadena, CA 91101}
\altaffiltext{6}{Kitt Peak National Observatory, National Optical Astronomy
Observatories, P.\,O. Box 26732, Tucson, AZ 85726}

\begin{abstract}
We compare the peculiar velocities measured
in the SBF Survey of Galaxy Distances with the predictions from
the density fields of the \iras\ 1.2~Jy flux-limited redshift survey and
the Optical Redshift Survey (ORS) to derive simultaneous
constraints on the Hubble constant $H_0$ and the density parameter
$\beta = \Omega^{0.6}/b$, where $b$ is the linear bias.  We find
$\bi=0.42^{+0.10}_{-0.06}$ and $\bo=0.26\pm0.08$ for the \iras\ and ORS
comparisons, respectively, and $H_0=74\pm4$ \kmsM\ (with an additional
9\% uncertainty due to the Cepheids themselves).  The match between
predicted and observed peculiar velocities is good for these values of
$H_0$ and $\beta$, and although there is covariance between the two
parameters, our results clearly point toward low-density cosmologies.
Thus, the unresolved discrepancy between the 
``velocity-velocity'' and ``density-density'' 
measurements of $\beta$ continues.
\end{abstract}


\section{Introduction}

Gravitational instability theory posits that the present-day
peculiar velocity $v_p$ of a galaxy
will equal the time-integral of the gravitational acceleration
due to nearby mass concentrations, i.e., 
$v_p = \langle g\rangle t$.  For galaxies outside 
virialized structures, linear perturbation theory can
be used to relate $v_p$ to the present-day local mass density:
\begin{equation}
	v_p(\mbox{\boldmath $r$}) \approx \frac{\Omega^{0.6} H_0}{4 \pi}
  \int{d^3\mbox{\boldmath $r^\prime$}\delta_m(\mbox{\boldmath $r^\prime$})
   \frac{\mbox{\boldmath $r^\prime - r$}}{|\mbox{\boldmath 
  $r^\prime - r$}|^3}}
\end{equation}
(Peebles 1980),
where $\Omega$ is the mass density of the universe in units
of the critical density, \h\ is the Hubble constant, and
$\delta_m(r)$ is the mass density fluctuation field.
A further common simplification is the linear biasing
model, $\delta_g = b\,\delta_m$, where $\delta_g$ is the fluctuation
field of the observed galaxy distribution, and $b$ is the linear bias.
With these assumptions, the observed peculiar velocity will
be proportional to the quantity $\beta = \Omega^{0.6}/b$.

Redshift surveys of complete samples of galaxies can be used to
determine the galaxy density in redshift space $\delta_g(z)$,
which is then smoothed and used to
predict the peculiar velocity field.  In this case, 
the distances are measured in units of the Hubble velocity and
$H_0$ cancels out of Eq.~(1).
Comparison with observed peculiar velocities from distance surveys
tied to the Hubble flow then yields a value for $\beta$ (see
Strauss \& Willick [1995] for a comprehensive review of the methods).
Recent applications using Tully-Fisher distances 
and the density field of the 
\iras\ 1.2~Jy redshift survey (Fisher \etal\ 1995)
have found best-fit values $\bi = 0.4$--0.6
(Schlegel 1995; Davis \etal\ 1996; da\,Costa \etal\ 1998;
Willick \etal\ 1997; Willick \& Strauss 1998),
where the subscript denotes the \iras\ survey.
Riess \etal\ (1997) have done the comparison with
Type~Ia supernovae (SNIa) distances and find $\bi = 0.40\pm0.15$.
On the other hand, the {\sc Potent} method (Dekel \etal\ 1990),
based on the derivative of Eq.~(1) but incorporating nonlinear terms,
gives a \bi\ about twice this value, most recently
$\bi = 0.89\pm0.12$ (Sigad \etal\ 1998).

The surface brightness fluctuation (SBF) method (Tonry \& Schneider 1988)
of measuring early-type galaxy distances is new
to the field of cosmic flows.  A full review of the
method is given by Blakeslee \etal\ (1999).
The $I$-band SBF Survey (Tonry \etal\ 1997, hereafter SBF-I) 
gives distances to about 300 galaxies out to $\sim\,$4000~\kms.
Tonry \etal\ (1999, hereafter SBF-II) used these data to construct a
parametric flow model of the surveyed volume,
but as the analysis made no use of galaxy distribution information,
it could not constrain $\Omega$. 
It did, however, allow for a direct tie between the
Cepheid-calibrated SBF distances and the unperturbed Hubble flow,
and thus a value for \h.  
Previous \h\ estimates with SBF (e.g., Tonry 1991; SBF-I) relied
on far-field ties to methods such as Tully-Fisher, $D_n$--$\sigma$,
and SNIa, and thus incurred additional systematic uncertainty. 
Unfortunately, this ``direct tie'' of SBF to the Hubble flow 
was still not unambiguous.
For instance, Ferrarese \etal\ (1999, hereafter F99) used a subset of
the same SBF data but a different parametric flow model to derive 
an $H_0$ 10\% lower than in SBF-II.

The present work uses SBF Survey distances in an initial 
comparison to the peculiar velocity
predictions from redshift surveys and
derives simultaneous constraints on \h\ and $\beta$.
The comparison also provides 
tests of linear biasing, the SBF distances, and claimed bulk flows.

\section{Analysis}
\label{sec:analysis}

We compute the redshift-distance (\czd) relation 
in the direction of each sample galaxy using the method of 
Davis \& Nusser (1994), which performs a spherical harmonic solution
of the redshift-space version of the Poisson equation.
Computations are done for both the \iras\ 1.2\,Jy (Fisher \etal 1995)
and ``Optical Redshift Survey'' (Santiago \etal\ 1995) catalogs
as a function of their respective density parameters \bi\ and \bo.
The smoothing scale of the \czd\ predictions is typically $\sim\,$500\,\kms.

The predictions are then compared to the SBF observations using
the $\chi^2$ minimization approach adopted by Riess \etal\ (1997),
who assumed a fixed ``redshift error'' of $\sigv{\,=\,}200$ \kms.
This term includes uncertainty in the linear predictions,
unmodeled nonlinear motions, and
true velocity measurement error;
we also report results with $\sigv{\,=\,}150$ \kms.  
We use the same subset of galaxies as in SBF-II,
namely those with high-quality data, $(V{-}I)_0 > 0.9$, and not
extremely anomalous in their velocities (e.g., Cen-45);  we also omit
Local Group members.  The full sample then comprises 280 galaxies.
Unlike the case for the SNIa distances, we lack a secure external tie
to the Hubble flow, so we successively rescale the SBF distances
to different values of \h\ and repeat the minimization.

The Nusser-Davis method cannot reproduce the multivalued \czd\ zones
of clusters, but it can produce major stall regions in the \czd\
relation.  Since such regions will have high velocity dispersions,
this deficiency in the method may be overcome with allowance for extra
velocity error.  A significant fraction of SBF galaxies
reside in the Virgo and Fornax clusters.  We adopt the following three
approaches in dealing with these.  
``Trial~1'' uses individual galaxy velocities but allows extra variance
in quadrature for the clusters according to: 
$\sigma_{\rm cl}(r) = \sigma_0/\sqrt{1+(r/r_0)^2}$, where 
%
$\sigma_0 = 700\,(400)$ \kms\ and $r_0=2\,(1)$ Mpc are adopted
for Virgo (Fornax).  These spatial profiles are meant
to mimic the observed projected profiles and parameters from 
Fadda \etal\ (1996) and Girardi \etal\ (1998); at large-radius
$\sigma_{\rm cl}$ follows the infall velocity profile (e.g., SBF-II).
``Trial~2'' 
uses a fixed velocity error
but removes the virial dispersions by assigning galaxies their
group-averaged velocities.  The groups are defined in SBF-I: 
29 galaxies are grouped into Virgo, 27 into Fornax, 7 into Ursa~Major,
and all other groups have 2-6 members, with 34\% of the sample being ungrouped.
``Trial~3'' is similar to trial~2, but 72 sample galaxies within 10~Mpc
of Virgo (about 50\% larger than the zero-velocity radius 
found in SBF-II) and 30 within 5~Mpc of Fornax are removed,
reducing the sample by 36\% to 178 galaxies. 

For each of the trials, we calculate $\chi^2$ for the comparisons
with both the \iras\ and ORS gravity fields, using \sigv\ of both
150 and 200 \kms, for a range in \h\ in steps of 0.1~\kmsM\
and a range in $\beta$ in steps of 0.1.  We then interpolate
in $\beta$  by cubic splines to find the \h-$\beta$ combination
minimizing $\chi^2$.

\section{Results}

Figure~\ref{fig:cont} shows the 68\%, 90\%, and 99\% joint confidence
contours on \h\ and $\beta$ from the $\chi^2$ analyses for trial~2 
(which gives the median best-fit \h) using $\sigv=200$ \kms.
Significant covariance exists between the two parameters.
Table~1 shows the $\chi^2$ minimization results for all the
different comparison runs. 
In order to give a more realistic impression of the true uncertainties,
the tabulated errors are for $\delta\chi^2 = 2.3$, 
the 68\% joint confidence range on 2 parameters,
except for the last column which 
varies \h\ within uncertainty limits to explicitly take account
of the covariance in $\beta$  (see below).

\begin{inlinefigure}\bigskip\medskip
\centerline{\includegraphics[width=0.90\linewidth]{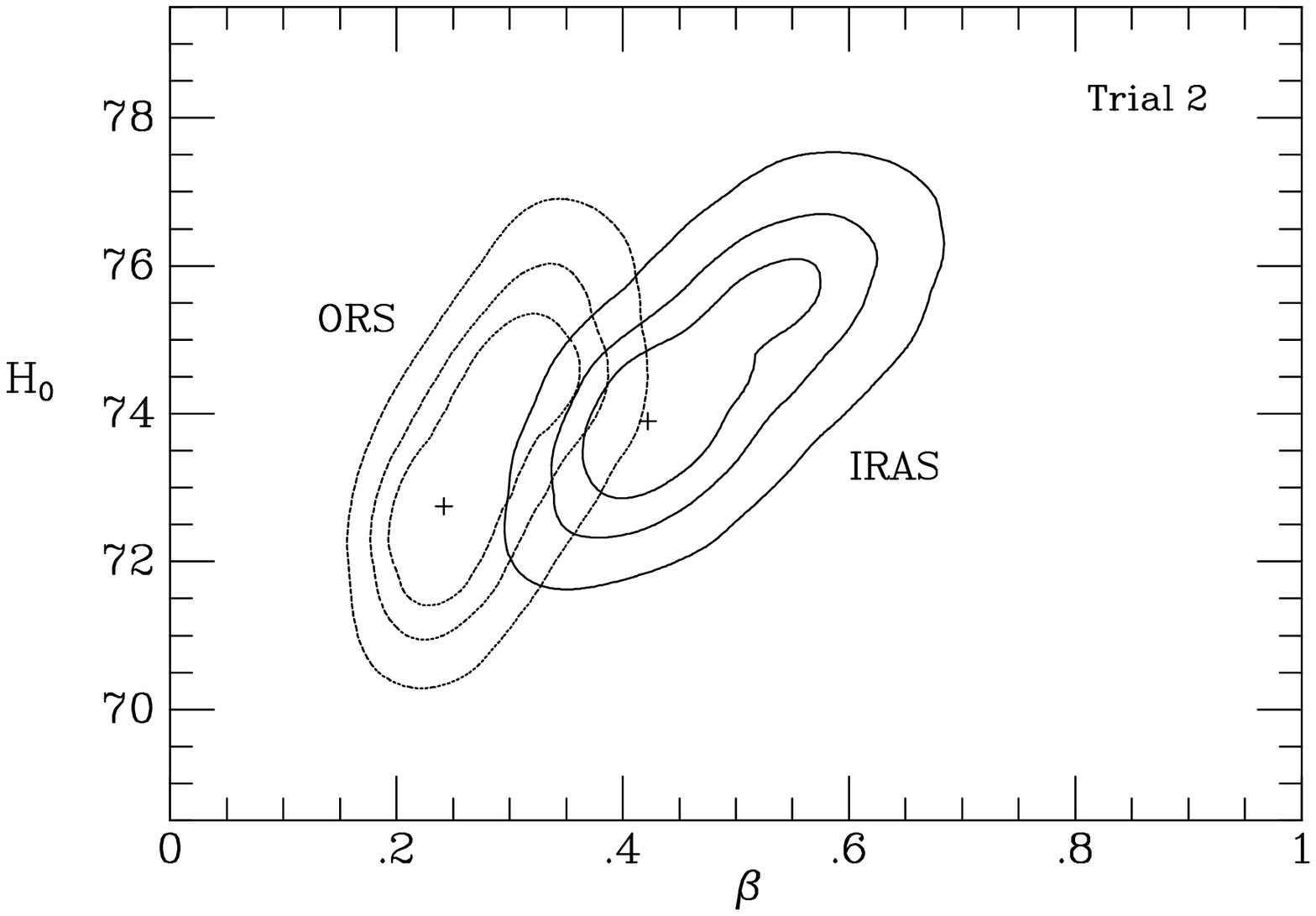}}
\caption{The 68\%, 90\%, and 99\% joint probability contours on $H_0$ 
and $\beta{\,\equiv\,}\Omega^{0.6}/b$ are plotted for the
``Trial~2'' comparison (see \S\ref{sec:analysis} for description)
of the SBF peculiar velocities against the 
\iras\ (solid curves at right) and 
ORS (dotted curves at left) density field predictions.
Crosses mark the best-fit positions.
\label{fig:cont}}
\end{inlinefigure}
\smallskip

\begin{inlinetable}\tablenum{1}
\begin{center}
\footnotesize
\caption{Results from Peculiar Velocity Comparisons\label{tab:egc}}
\newdimen\digitwidth \setbox0=\hbox{\rm0} \digitwidth=\wd0 \catcode`?=\active
\def?{\kern\digitwidth}
\begin{tabular}{p{1.32cm}cccccc}
\tableline 
\tableline \tablevspace{2pt}
Run & ~\sigv & $\chi^2_\nu$ & $\beta$~??~$\pm$ &
~$H_0$??$\pm$ & $\beta_{H_0{=}74}\,\pm$? \\ \tablevspace{2pt}
\hline \tablevspace{3pt}
 IRAS--1\dotfill &  200 &  0.99 &  0.39~~0.06 & 73.2~~1.2 & 0.42~~$^{+0.10}_{-0.07}$ \\ \tablevspace{3pt}
 IRAS--1\dotfill &  150 &  1.21 &  0.40~~0.07 & 73.7~~1.1 & 0.41~~$^{+0.09}_{-0.06}$ \\ \tablevspace{3pt}
 IRAS--2\dotfill &  200 &  0.89 &  0.43~~0.06 & 73.9~~1.0 & 0.43~~$^{+0.10}_{-0.06}$ \\ \tablevspace{3pt}
 IRAS--2\dotfill &  150 &  1.18 &  0.45~~0.06 & 74.3~~1.0 & 0.44~~$^{+0.08}_{-0.05}$ \\ \tablevspace{3pt}
 IRAS--3\dotfill &  200 &  1.05 &  0.40~~0.08 & 74.1~~1.2 & 0.40~~$^{+0.10}_{-0.05}$ \\ \tablevspace{3pt}
 IRAS--3\dotfill &  150 &  1.33 &  0.41~~0.07 & 74.4~~1.2 & 0.41~~$^{+0.07}_{-0.05}$ \\ \tablevspace{2pt}
\tableline\tablevspace{3pt}
  ORS--1\dotfill &  200 &  1.05 &  0.23~~0.05 & 72.1~~1.2 & 0.25~~$^{+0.07}_{-0.05}$ \\ \tablevspace{3pt}
  ORS--1\dotfill &  150 &  1.28 &  0.24~~0.05 & 72.4~~1.0 & 0.27~~$^{+0.06}_{-0.05}$ \\ \tablevspace{3pt}
  ORS--2\dotfill &  200 &  0.97 &  0.25~~0.05 & 72.8~~1.3 & 0.29~~$^{+0.06}_{-0.07}$ \\ \tablevspace{3pt}
  ORS--2\dotfill &  150 &  1.29 &  0.27~~0.05 & 73.1~~1.2 & 0.30~~$^{+0.05}_{-0.05}$ \\ \tablevspace{3pt}
  ORS--3\dotfill &  200 &  1.12 &  0.22~~0.05 & 74.1~~1.4 & 0.21~~$^{+0.05}_{-0.04}$ \\ \tablevspace{3pt}
  ORS--3\dotfill &  150 &  1.43 &  0.24~~0.06 & 74.6~~1.5 & 0.22~~$^{+0.05}_{-0.04}$ \\ \tablevspace{2pt}
\tableline
\tablevspace{4pt}\multicolumn{6}{l}{Columns list:
the comparison run (designated as survey--trial); 
}\\ \multicolumn{6}{l}{assumed\, small-scale\, velocity\, error\, \sigv\ (\kms);\,
reduced $\chi^2$ 
}\\ \multicolumn{6}{l}{(degrees of freedom\, =\, number of galaxies minus 2);\,
best-fit 
}\\ \multicolumn{6}{l}{$\beta$ and $H_0$ (\kmsM);
and  best-fit $\beta$ for
$H_0 = 74\pm1.4$.}
\\
\end{tabular}
\end{center}
\end{inlinetable}
\medskip\medskip

On average, the \iras\ comparisons prefer
$\bi\approx0.4$ and $H_0\approx74$, while the ORS,
which more densely samples the clusters, prefers
$\bo\approx0.25$ and $H_0\approx73$.  Of course, there can be
only one value of \h\ for the sample galaxies.
The table shows that when the cluster galaxies are
removed in trial~3, the best-fit \h\ increases to 74 for the ORS,
while it changes little for \iras. 
We adopt $H_0 = 74\pm1.4$ as the
likely value from this velocity analysis,
where we use the median \h\ error for the \iras\ trials combined
in quadrature with the variance in \h\ among the trials.
The error increases to $\pm4$~\kmsM\ 
when the 5\% statistical uncertainty in the tie to the
Cepheid distance scale is added in quadrature (SBF-II).
Finally, the estimated uncertainty in the Cepheid scale itself is
$\sim\,$9\% (F99; Mould \etal\ 1999).
These uncertainties in \h\ due to the distance zero point
have no effect on the uncertainty in $\beta$.

\begin{inlinefigure}\bigskip 
\centerline{\includegraphics[width=0.90\linewidth]{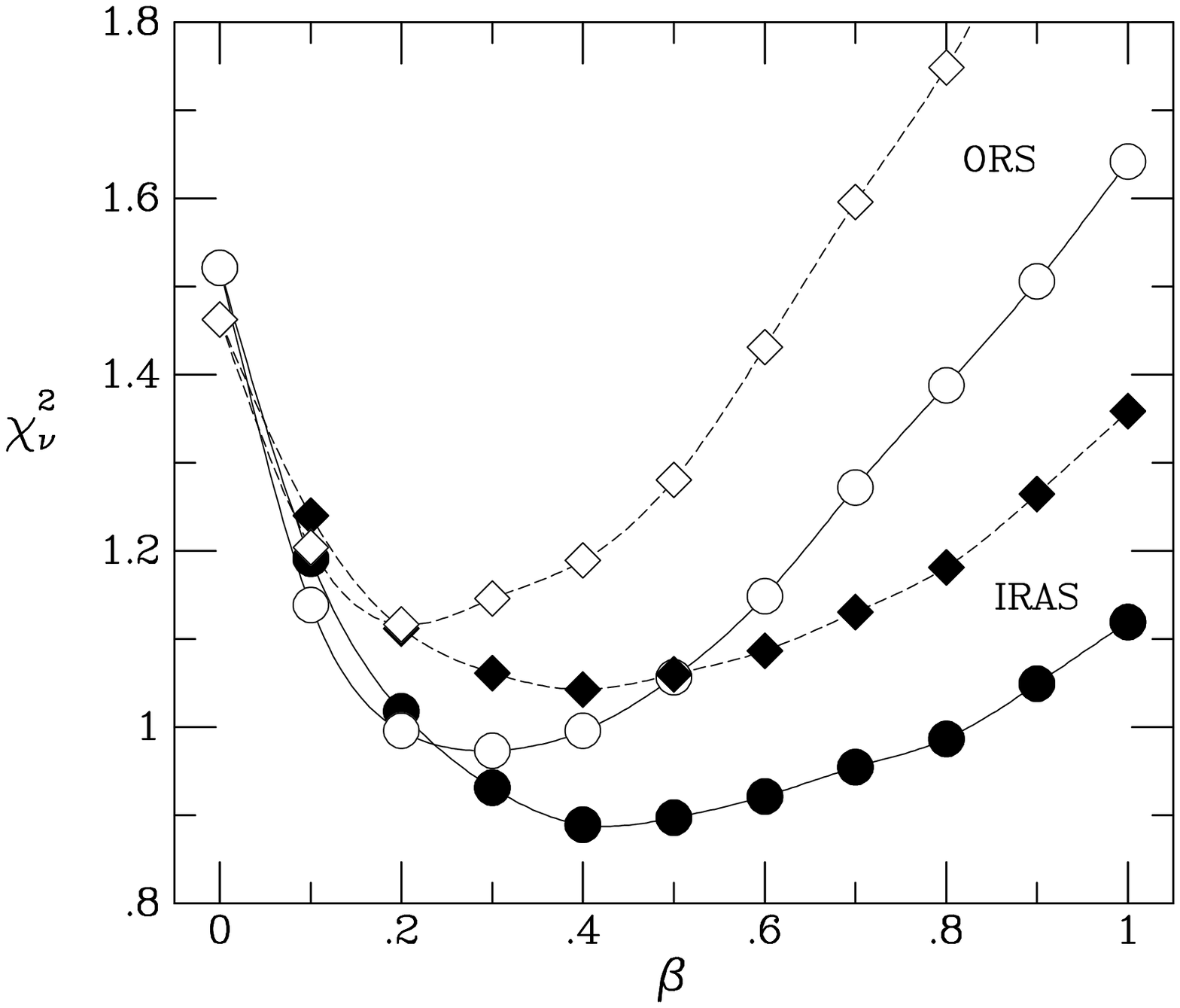}}\smallskip
\caption{The reduced $\chi^2$ is plotted as a function of 
$\beta$ for $H_0{\,=\,}74$ \kmsM\ and $\sigv = 200$ \kms. 
Circles connected by solid spline curves show results from
``Trial~2,'' which includes all 280 galaxies,
and diamonds connected by dashed curves show 
results from ``Trial~3,'' which removes
102 galaxies near the Virgo and Fornax clusters.
Solid/open symbols are used for the \iras/ORS comparisons.
\label{fig:chi}}
\end{inlinefigure}

Figure~\ref{fig:chi} illustrates how the reduced $\chi^2$
for $H_0=74$ and $\sigv=200$ varies with $\beta$
for trials~2 and~3 (spanning the range in best-fit $\beta$
and $\chi^2_\nu$) of the \iras\ and ORS comparisons.  
Overall, the last columns of Table~1 indicate $\bi=0.42^{+0.10}_{-0.06}$
and $\bo=0.26\pm0.08$, where
the errors include the uncertainty for a given trial (including the
2\% uncertainty in \h\ from the velocity tie) and the
variation among the trials.
The results are independent of whether one
takes a median or average of the trials, or adopts either
$\sigv=150$ or 200 \kms. 
However, \bi\ and \bo\ would increase for example by $\sim\,$30\%
to 0.56 and 0.33, respectively, if \h\ were known
to be 5\% larger {\it for a fixed distance zero point},
or decrease by 15--20\% if \h\ were 5\% smaller,
but again, the $\chi^2$ analysis indicates \h\ is constrained to 2\% 
for a fixed distance zero~point.

Finally, Figure~\ref{fig:aitoff} compares the 
$\bi = 0.4$ predicted and $H_0 = 74$ observed peculiar
velocities in the Local Group frame for all the galaxies,
using their group-averaged velocities to limit noise.
Given that the two sets of peculiar velocities were derived
independently (apart from our adjustment of \h\ and \bi),
the agreement is quite good.
The predictions resemble a heavily smoothed version of 
the observations, and $\chi^2_\nu$ confirms this,
although the residuals may hint at a slightly misaligned dipole.
Large negative residuals near $(l,b)\sim(283^\circ,+74^\circ)$
indicate Virgo backside infall in excess of
what the spherical harmonic solution can produce.
These issues will be addressed in more detail by
a forthcoming paper (Willick \etal\ 2000, in preparation).

\begin{inlinefigure}
\centerline{\includegraphics[width=0.90\linewidth]{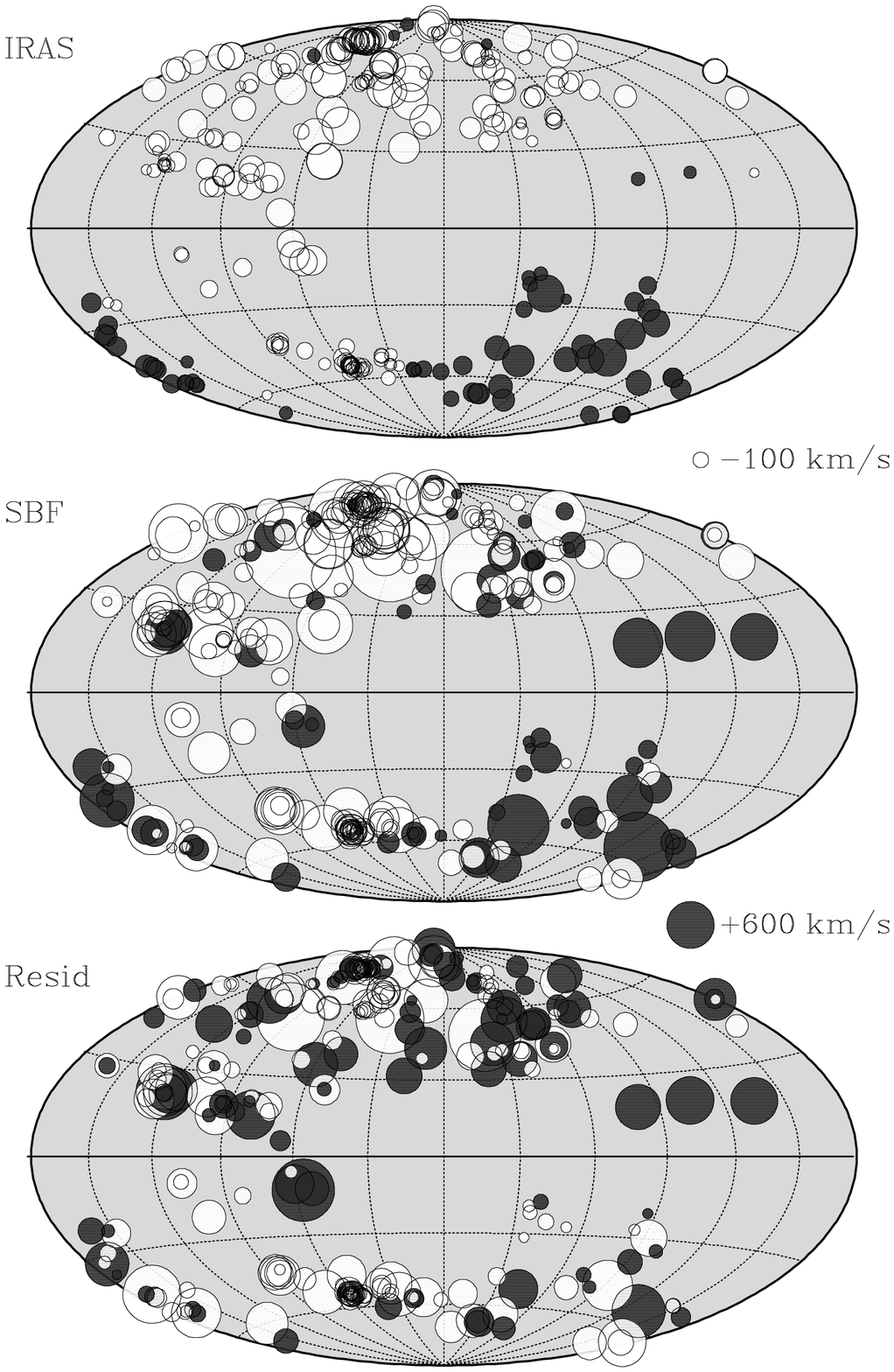}}\smallskip
\caption{Predicted peculiar velocities from the \iras\ 
survey (top) with $\bi=0.4$, observed peculiar velocities
from the SBF survey (middle) with $H_0=74$ \kmsM,
and the residuals (observed minus predicted, bottom) are shown
in Galactic coordinates and the Local Group frame. 
Light/dark circles show negative/positive peculiar velocities,
coded in size as shown; smaller symbols are plotted on
top of larger ones for clarity.
Unlike the predictions, the observations have no smoothing
applied and contain both distance and velocity errors,
causing the greater range in symbol size.
\label{fig:aitoff}}
\end{inlinefigure}

\section{Discussion}

We have found the most consistent results for $H_0 = 74$ \kmsM,
intermediate between the favored values of 
77 and 71 reported by SBF-II and the $H_0$~Key Project's
analysis of SBF (F99), respectively. The differences 
in these ``SBF \h'' values
result entirely from the flow models\footnote[1]{We have
changed the Key Project value by 1.6\% to be appropriate for our 
distance zero~point, since our concern here is with the tie 
to the Hubble flow.}.
SBF-II explored a number of parametric flow models
ranging from a pure Hubble flow plus dipole to models
with two massive attractors and a residual quadrupole,
as well as some with a local void component.
The pure Hubble flow model gave the same \h\ as obtained by F99.
Adding Virgo and Great Attractors gave $H_0{\,=\,}73.5$,
and the additional Local Group-centered quadrupole gave
a further 6\% increase in \h.
Each component significantly improved the model likelihood, but
it was suggested that the quadrupole arose from inadequate modeling
of the flattened Virgo potential.  If so, then \h\ may be better
estimated using a sharply cutoff, Virgo-centered quadrupole.
Figure~23 of SBF-II shows that such a model
would indeed yield $H_0 \approx 74$.


We also note that most of the SBF-II models had a significant excess 
Local Group peculiar motion of $\sim\,$190\,\kms,
unlike the \iras\ density field predictions (e.g., Willick etal\ 1997).
However, when the excess was modeled
as a ``push'' from a large nearby void, \h\
dropped from 78 to 73 because of the underdensity introduced.
This model achieved the best likelihood of any considered,
but the treatment of the void was deemed too ad~hoc
to qualify as a standard component of the flow model.
Thus, it may be that the SBF-II result for \h\ suffered from
the arbitrariness inherent in parametric modeling.
However, the good match we find between the predicted and observed
velocity fields supports the claim by SBF-II that,
after accounting for the attractor infalls,
any bulk flow
of the volume $c{z}\lta3000$ \kms\ is 
$\lta200$ \kms, as the structure within this volume accounts
for most of our motion in the cosmic microwave background rest frame
(see Nusser \& Davis 1994).

The best-fit values of $\bi=0.42^{+0.10}_{-0.06}$
and $\bo=0.26\pm0.08$ are similar to the
$\bi=0.40\pm0.15$ and $\bo=0.3\pm0.1$ results found by Riess \etal\ (1997)
using 24 SNIa distances out to $c{z}\sim9000$ \kms, nearly 3~times
our survey limit.  It is interesting that despite the close
agreement on $\beta$, SBF and SNIa still disagree by nearly 10\% on \h\
(e.g., Gibson \etal\ 1999).  This implies that the discrepancy is mainly
due to the respective distance calibrations against the Cepheids
(otherwise the \h\ offset would cause major disagreement on $\beta$).
We also note that the ratio $\bi/\bo \approx 1.6$ agrees well 
with the estimates by Baker \etal\ (1998).

Our results are consistent with all other recent comparisons of the gravity
and velocity fields (so-called ``velocity-velocity'' comparisons),
regardless of the distance survey used.  Thus, the near factor-of-two
discrepancy with the {\sc Potent} analysis (a ``density-density''
comparison) of Sigad \etal\ (1998) persists.  Since the latter analysis
is done at larger smoothing scales, one obvious explanation
is non-trivial, scale-dependent biasing, and some simulations may 
give the needed factor-of-two change in bias over the relevant range
of scales (Kauffmann \etal\ 1997; but see Jenkins \etal\ 1998).  However, 
at this point, the inconsistency in the results of the two types of
analysis remains unexplained.


We are currently in the process of analyzing the SBF data set 
using methods that deal directly with multivalued redshift zones 
and incorporate corrections to linear gravitational instability
theory to take advantage of SBF's ability to probe the small-scale
nonlinear regime (Willick \etal\ 2000, in preparation).
We also plan in the near future to use the data for 
a ``density-density'' determination of $\beta$.
In addition, we are working towards a direct, unambiguous
tie of SBF to the far-field Hubble flow, which will 
significantly reduce the systematic uncertainty in $\beta$.

\medskip

\acknowledgments
We thank Laura Ferrarese for sharing Key results
prior to publication.
This work was supported by NSF grants AST9401519
and AST9528340.
JPB thanks the Sherman Fairchild Foundation for support.
\hfil\\

\end{document}